# SuperCoder2.0: Technical Report on Exploring the feasibility of LLMs as Autonomous Programmer

Anmol Gautam
*SuperAGI Research*
Bengaluru, India

Kishore Kumar
*SuperAGI Research*
Bengaluru, India

Adarsh Jha
*SuperAGI Research*
Bengaluru, India

Mukunda NS
*SuperAGI Research*
Bengaluru, India

Ishaan Bhola
*SuperAGI Research*
Bengaluru, India

*Abstract*—We present SuperCoder2.0, an advanced autonomous system designed to enhance software development through artificial intelligence. The system combines an AI- native development approach with intelligent agents to enable fully autonomous coding. Key focus areas include a retry mechanism with error output traceback, comprehensive code rewriting and replacement using Abstract Syntax Tree (ast) parsing to minimize linting issues, code embedding technique for retrieval-augmented generation, and a focus on localizing methods for problem-solving rather than identifying specific line numbers. The methodology employs a three-step hierarchical search space reduction approach for code base navigation and bug localization:utilizing Retrieval Augmented Generation (RAG) and a Repository File Level Map to identify candidate files, (2) narrowing down to the most relevant files using a File Level Schematic Map, and (3) extracting 'relevant locations' within these files. Code editing is performed through a two-part module comprising CodeGeneration and CodeEditing, which generates multiple solutions at different temperature values and replaces entire methods or classes to maintain code integrity. A feedback loop executes repository-level test cases to validate and refine solutions. Experiments conducted on the SWE-bench Lite dataset demonstrate SuperCoder2.0's effectiveness, achieving correct file localization in 84.33% of cases within the top 5 candidates and successfully resolving 34% of test instances. This performance places SuperCoder2.0 globally on the SWE-bench leaderboard. The system's ability to handle diverse repositories and problem types highlights its potential as a versatile tool for autonomous software development. Future work will focus on refining the code editing process and exploring advanced embedding models for improved natural language to code mapping.

## I. INTRODUCTION

SuperCoder2.0 is an advanced autonomous system designed to enhance software development through artificial intelligence. Combining an AI-native development platform with intelligent agents, it allows for fully autonomous coding, particularly in Python. This system introduces several innovative features such as an advanced retry mechanism with error output traceback, comprehensive code rewriting to avoid linting issues, a unique code embedding technique for retrieval-augmented generation, and a focus on localising methods for problem-solving rather than identifying specific line numbers. Current Landscape of Autonomous Systems in Software Development The domain of AI-assisted development has seen substantial research and development. Techniques such as automated code generation, error detection, and project management have demonstrated significant improvements in developer productivity and code quality. Research on AI-native platforms emphasizes the importance of integrating AI models into the development lifecycle to enhance deployment fluidity and operational efficiency. In this context, autonomous systems like OpenAI's Codex [1], which powers GitHub Copilot, exemplify AI's potential to transform traditional coding practices by enabling more intuitive coding environments and reducing manual efforts. Techniques such as RAG [2], combining strengths of retrieval-based and generation-based models, are pivotal in enhancing contextually accurate code suggestions.

SuperCoder2.0 incorporates advanced error handling and retry mechanisms, critical for autonomous systems. These features minimize debugging time and enhance robustness through proactive error management, automated retries, and error output traceback. The implementation of error output traceback and automated retries in SuperCoder2.0 aligns with advanced techniques seen in recent studies, which highlight the importance of learning from previous mistakes to improve AI model performance. Research demonstrates that proper error handling can significantly reduce manual debugging efforts and improve overall system reliability [3] [4]. The comprehensive approach of code rewriting of SuperCoder2.0 involves rewriting the entire files to avoid linting issues, a method that ensures a consistent and cohesive codebase. This technique is crucial for maintaining code quality and avoiding fragmented, inconsistent updates, a challenge frequently noted in modern software development practices. Leveraging RAG, particularly with specialized code embeddings, SuperCoder2.0 enhances its ability to provide relevant and accurate code suggestions. This is especially pertinent given the growing body of research advocating for improved accuracy and relevance in AI-driven code assistance tools. The embedding of code documents using the code embeddings instead of traditional embeddings, advances beyond standard RAG approaches by offering more contextually relevant retrievals with natural language input. Localisation of Problem-Solving is another place where SuperCoder2.0 takes a unique approach to avoid identifying a wrong location. Traditional debugging approaches often focus on identifying specific line numbers where issues occur. However, the localisation of methods rather than specific lines, as practised by SuperCoder2.0, aligns with current research that suggests higher abstraction methods can improve the understanding and solving of complex coding problems [5]. This method provides a more strategic approach to debugging and problem resolution, making it easier to manage and fix intricate issues that span multiple code segments.

## II. Literature Survey

The rapid advancement in artificial intelligence, particularly in large language models (LLMs), has improved the development of autonomous systems capable of performing complex tasks managed by humans. This is especially true in the domain of software development, where the integration of AI-driven autonomous agents is increasingly adopted. Autonomous Software Development Systems are evolving to automate various phases of the software lifecycle, from requirements gathering and design to coding, testing, and maintenance. One such benchmark, SWE-bench [6], has emerged as a pivotal tool in assessing the capabilities of language models in performing software engineering tasks autonomously. SWE-bench is designed to measure the effectiveness of language models in handling tasks that range from basic code comprehension to more complex activities such as code synthesis and bug fixing. The benchmark provides a structured evaluation environment, offering a diverse set of challenges that reflect real-world software development scenarios. SWE-bench has been instrumental in highlighting the strengths and limitations of various autonomous systems. For instance, it has been used to compare the performance of different LMs in coding tasks, revealing significant variations in their ability to understand and generate syntactically correct and semantically meaningful code. SWE-bench has played a crucial role in driving innovation in the development of more sophisticated LMs. Providing a clear set of performance metrics encourages researchers and developers to push the boundaries of what these models can achieve. The benchmark has been vital in identifying areas where current models fall short, such as in handling highly complex or creative tasks that require deep contextual understanding. LLMs like GPT-4 by OpenAI and Claude by Anthropic are pivotal in advancing autonomous software development. Initially, LLMs were designed to handle basic natural language processing tasks, including translation and summarization. Their potential in code generation became evident with early models like GPT-2, which could generate simple [7]. Over time, models like GPT-4 have shown marked improvements in both the accuracy and complexity of generated code, evolving from handling basic tasks to performing sophisticated programming activities [8]. GPT-4 builds on its predecessors' architectures with enhanced capabilities, trained on vast and diverse datasets, enabling it to understand complex programming languages and generate more sophisticated code. Its ability to interpret nuanced queries and produce functional code makes it a powerful tool for automating software development tasks [9].

Claude, with its focus on safety and alignment, has developed robust mechanisms to generate precise, error-free code, ensuring higher reliability in autonomous coding systems [10]. The emphasis on reducing biases and achieving high fidelity in generated code has made Claude particularly beneficial for applications that require safe and reliable outputs. Through iterative training on extensive and varied datasets, Claude ensures the correct semantic interpretation of programming tasks, thus generating precise and dependable code [11]. These models demonstrate the potential to automate various aspects of software development, providing a technical backbone for more sophisticated autonomous coding systems.

The incorporation of LLMs into autonomous agents represents a significant leap toward realizing fully autonomous software development systems. Frameworks like LangChain enable LLMs to function as part of autonomous agents capable of executing complex tasks with minimal human intervention [12]. These agents can request and utilize information, develop strategies, and make decisions based on the extensive knowledge embedded in the LLMs, significantly enhancing their capabilities in autonomous coding environments. Methodologies like Reasoning and Acting (ReACT) and Chain-of-Thought (COT) further enhance the application of LLMs.

ReACT integrates reasoning with action, enabling LLMs to autonomously devise and execute plans, crucial in real-time coding environments [13]. This approach leverages the model's reasoning capabilities to interpret tasks and generate corresponding actions autonomously. By integrating structured reasoning and action-based planning, ReACT empowers LLMs to handle complex coding tasks with greater accuracy and consistency [14]. COT further improves the problem-solving capabilities of LLMs by breaking down tasks into sequential steps, enhancing their ability to generate coherent and functional code [15]. This step-by-step reasoning approach allows models to process and address each part methodically, making it particularly effective in software development where logical sequences and dependencies are critical. Studies have revealed that COT prompting enhances an LLM's capacity to generate clear and coherent code by improving its logical reasoning processes [16].

The capability of LLMs to generate code has been extensively researched, demonstrating the profound impact these models can have on software development. Benchmark datasets like CodeXGLUE evaluate and advance LLM capabilities in code-related tasks, highlighting how pre-trained models can be fine-tuned for specific applications [17]. This benchmark includes a suite of tasks designed to test various aspects of code intelligence, from code completion to translation between different programming languages. Research efforts have developed tools like GitHub Copilot, built on advanced LLMs like Codex, a descendant of GPT-3, which have practical applications in improving developer productivity by providing context-aware code completions. Copilot can generate boilerplate code, suggest entire functions, and even automate repetitive coding tasks, significantly reducing the manual workload for developers [18].

Another empirical study delves into the adaptive capabilities of models like GPT-4 across various programming languages and tasks. Fine-tuning these models on domain-specific datasets has shown to lead to high accuracy and functionality in code generation, underscoring the versatility and robustness of LLMs in different coding environments [19]. Further research has explored the integration of LLMs in more complex programming tasks, demonstrating how models trained on diverse code examples can understand and replicate intricate programming patterns. This approach significantly enhances the model's ability to generate contextually appropriate and syntactically correct code that adheres to best practices [20]. Multi-modal code generation, which integrates natural language tasks with coding tasks within a single model framework, has demonstrated improved code generation capabilities, achieving better contextual understanding and producing more accurate and coherent code [21].

Numerous research efforts and advancements in the AI programmer space have led to the development of sophisticated tools like OpenDevin and Agentless. OpenDevin facilitates end-to-end programming tasks, bridging the gap between natural language specifications and functional software [22]. This tool leverages state-of-the-art LLMs to understand complex programming requirements and produce accurate, high-quality code, effectively automating comprehensive development processes. The introduction of such tools showcases the potential for AI systems to handle substantial portions of the software development lifecycle.

Agentless, another cutting-edge AI tool, exemplifies current strides in autonomous software engineering by focusing on intelligent automation for maintaining and managing codebases. It employs an initial file localization and code generation phase that reduces complexity and leverages the capabilities of LLMs [23]. Agentless utilizes advanced reasoning capabilities and machine learning algorithms to understand code dependencies, suggest improvements, and perform refactoring, thus enhancing both the efficiency and reliability of software projects. AutoCodeRover enhances software code quality through a combination of static and dynamic analysis techniques, autonomously navigating through the codebase and applying transformations to improve performance and readability [24].

The CodeAct agent consolidates an LLM agent's actions into a unified code action space, operating in a turn-based environment where it can converse with the user or perform actions such as running scripts and navigating files. This approach allows the LLM to reason, create plans, and incorporate feedback from the user, improving robustness [25]. A multi-agent approach for autonomous software development is presented where different real-world roles involved in the software development lifecycle are assigned to multiple agents, which communicate to solve problems collaboratively. This task graph methodology allows for efficient planning and execution of tasks by different agents [26].

The agent-based approaches have their drawbacks, as discussed in research supporting the agentless paradigm. The agentless approach employs a simplified initial file localization and a code generation phase, bypassing the need for complex agent-based tools and utilizing repo maps for file searches, significantly reducing complexity [27]. The modular architecture for software-engineering AI agents (MASAI) follows a similar strategy, where different LLM-powered sub-agents are instantiated with specific objectives like edit localization, fault fixing, and issue reproduction, distributing responsibilities and simplifying problem-solving [28].

Aider utilizes static analysis to offer a concise overview of repositories, identifying files that need editing and performing iterative modifications until the code is syntactically correct and passes existing tests [29]. Another approach, Moatless, employs a semantic search tool to use natural language queries to find relevant code segments within the repository, simplifying the localization and correction of code [30].

In conclusion, the fusion of LLMs like GPT-4 and Claude with methodologies such as ReACT and COT, coupled with innovative tools like OpenDevIN and Agentless, signifies a transformative period in autonomous software development. By automating various programming activities, these AI systems promise to enhance productivity, reduce errors, and allow human developers to focus on more strategic and creative aspects of software engineering. As research continues to evolve, the integration of AI in the programming field is set to revolutionize the development, maintenance, and optimization of software, paving the way for a new era of autonomous coding.

III. METHODOLOGY

This section elucidates the operational framework of SuperCoder2.0, focusing on its approach to codebase navigation and issue resolution. The methodology addresses two critical aspects of automated debugging: [33] efficient codebase traversal with precise localization of the target code section, and [34] generation and implementation of corrective code. While individual components of this approach have precedents in existing literature, the novel integration and sequential application of these elements yield significant performance improvements. The methodology is bifurcated into two primary phases: Search and Edit. The Search phase employs a hierarchical approach to narrow down the problem space, while the Edit phase utilizes advanced code generation and modification techniques. This structured approach enables SuperCoder2.0 to effectively navigate complex codebases and implement targeted solutions, demonstrating enhanced efficiency in autonomous software development tasks. The system leverages state-of-the-art language models, specifically GPT-4 and Claude Sonnet 3.5, for code generation and analysis tasks, while Jina Code Embeddings are utilized for efficient code retrieval and semantic search operations. This combina-tion of advanced models contributes to the system's robust performance across diverse programming challenges.

*A. Search*

To effectively resolve issues within a codebase, it is crucial to precisely identify the root of the problem, henceforth referred to as the 'relevant location'. In SuperCoder2.0, we define the 'relevant location' as belonging to one of three hierarchical levels: Method/Function, Class, or Top-Level. Top-level code encompasses elements outside any class or function, including import statements. Class-level code is contained within a class but external to its methods, while method/function-level code is encapsulated within specific methods or functions. The primary objective of the Search module is to localize the bug's residence. When identified within a method or function, the module returns the method/function name along with its start line number, facilitating disambiguation between identically named methods. Empirical observations on the SWE-Bench-Lite dataset demonstrated that accurate method localization yielded optimal results. Considering that most LLMs can generate up to 4,000 tokens in a single iteration, we posit that methods or functions adhering to best software engineering practices are likely to fall within these generation limits. Our approach implements a hierarchical search space reduction strategy, systematically narrowing down from file-level to method-level granularity. This process is executed in three distinct steps:

- Identification of candidate files from the repository, utilizing RAG and a Repository File Level Map.
- Refinement of candidate files to the most relevant subset, employing a File Level Schematic Map.
- Extraction of 'relevant locations' within the identified files.

Each step in this process involves a discrete LLM call. Notably, our empirical findings suggest that providing comprehensive context and posing targeted questions in a single LLM call proves more effective than an iterative, reason-action-observation based agentic approach. This observation aligns with findings from the Agentless team. The subsequent sections will elaborate on each step of this process in detail.

*1) RAG and Repository File Level Map:* The initial phase involves the creation of a vector store for the codebase, facil- itating the retrieval of files most pertinent to a given problem statement. This process remains largely static for a given codebase. The document structure employed for embedding and searching is detailed in Listing 1.

Listing 1. Embedding document structure with metadata

```
{
    "document": "{
        Method {{methodname}} with
        arguments {{args}} have signature as
        {{signature}} is described
        using {{docstring}} also
        have {{decorators}} as decorators
        and return statement described
        as {{return statement}}.
    },
    "file name":<<name of the file>>,
    "parent class":<<name of the parent class or None if its a method>>
}"
```

We implemented method-level embeddings to identify the most relevant methods for a given problem statement. Each method or function within the codebase was parsed and its information was converted into a string representation for embedding generation. The embedding incorporates the method or function name, signature, return statements, and docstring. These individual document representations were then vectorized and stored in the vector store. The Repository Level File Map was generated through a recursive parsing of the codebase, extracting relative filenames from the root folder. This map serves as an additional resource for inferring the most relevant files. A snippet of this is shown in Listing 2.

Listing 2. Repository Level File Map

```
{
    "matplotlib": ["
        setupext.
        py","setup.
        py",
        "tests.py"
    ],
    "matplotlib/tools": ["
        gh api.py",
        "boilerplate.py",
        "cache zenodo svg.py","
        github stats.py",
        "embed js.py",
        "triage tests.
        py","run examples.
        py","subset.py",
        "visualize tests.py","
        memleak.py",
        "make icons.py"
    ],
    "matplotlib/ci": [
        "check wheel licenses.py"
    ],
    .
    .
    .
}
```

Utilizing these two distinct modules, we process a problem statement by converting it into 'N' queries via the Query-Generation module, which employs a separate LLM call. These 'N' queries are then used to interrogate the vector store, retrieving the most relevant filenames from the extracted chunks. Concurrently, we invoke the FileLocater module through another LLM call, which processes the Repository

File Level Map to return the top 'M' filenames. The final set of candidate files is determined by performing a union operation on these two lists of filenames. This dual-pronged approach, combining semantic search through RAG with structural analysis via the Repository File Level Map, enhances the robustness and accuracy of our file localization process. The integration of these complementary methods allows for a more comprehensive and nuanced identification of relevant code sections, thereby improving the overall efficiency of the bug localization process.

*2) File Prioritization using File Level Schematic Map:* Building upon the set of candidate files identified in the previ- ous step, this phase involves the creation of a schematic rep- resentation for each file, providing a comprehensive overview of its internal structure. This schematic representation en- compasses detailed descriptions of all classes and functions within the file. The schema includes class names, contained methods, method arguments, applied decorators, and available docstrings. This abstracted representation enables the Large Language Model (LLM) to gain a high-level perspective of the filegroup, facilitating more accurate identification of files closely related to the problem statement. The PreAssimilation Module is then employed to further refine the search space, narrowing down the potential bug location to the top 'L' files. Notably, the value of 'L' is dynamically determined by the LLM itself, allowing for adaptive file selection based on the complexity and nature of the problem. Through extensive experimentation, we observed that in exploratory scenarios, such as when SuperCoder2.0 is tasked with creating a new fea- ture, allowing the LLM to autonomously select relevant files yields optimal results. For the specific evaluation conducted on SWE Bench Lite, we implemented a constraint limiting 'L' to a maximum of two files. In this configuration, the PreAssimilator module returns either the single most relevant file or, in more complex cases, the two most pertinent files along with a rationale for their selection. This file and method localisation approach (Figure 1) strikes a balance between comprehensive search and computational efficiency, enabling SuperCoder2.0 to focus on the most promising areas of the codebase while maintaining the flexibility to handle diverse problem types. The integration of LLM-driven file selection with the structured schematic representation enhances the sys- tem's ability to navigate and understand complex codebases, thereby improving its overall problem-solving capabilities.

*3) Localization of 'Relevant Locations':* Following the identification of the top 'L' files in the preceding steps, the process advances to pinpointing specific areas within these files where modifications are required. This task is accomplished through the deployment of the CoderParser Module, a sophisticated component designed to analyze the problem statement in conjunction with the entire file content. The CoderParser Module performs a comprehensive analysis, outputting both the precise location and a detailed plan for necessary code alterations. The full specifications of this pro- cess are delineated in Appendix 1c. Our approach categorizes potential code changes into three distinct hierarchical levels:

- Top-Level code modifications
- Method or Function level alterations
- Class-Level adjustments

This stratified approach was developed in response to em- pirical observations revealing that line-level edits frequently resulted in linting issues, particularly in languages with strict indentation requirements such as Python. Consequently, we adopted a strategy of high-level code replacement, focusing on class-level or method-level modifications generated by the LLM. For top-level code changes, the LLM is instructed to specify start and end line numbers delineating the sec- tion requiring modification. This demarcated code segment is then replaced in its entirety. This methodology ensures the maintenance of code structure and integrity while allowing for comprehensive modifications. It mitigates potential syntax errors and maintains consistency with the existing codebase architecture. The specific implementation details of this editing process are elaborated upon in the subsequent Edit section. By employing this hierarchical, structure-aware approach to code modification, SuperCoder2.0 achieves a balance between gran- ular problem-solving and preservation of overall code quality and consistency. This method enhances the system's ability to implement complex changes while minimizing unintended side effects or structural disruptions to the codebase.

*B. Code Modification and Insertion*

The Code Modification and Insertion module is responsible for implementing the actual code changes as determined by the preceding analysis. This module operates based on the action plan provided by the CodeParser module, processing each plan element sequentially. The module receives relevant class or method information, or in the case of top-level code modifications, it obtains the code segment defined by pre- determined start and end lines. The module's architecture comprises two primary components:

- CodeGeneration Module
- CodeEditing Module

The CodeGeneration Module leverages advanced LLMs to generate new or modified code segments based on the provided action plan and context. In contrast, the CodeEditing Module operates independently of LLMs, utilizing the Abstract Syntax Tree (AST) library in Python for precise code manipulation. This module processes the class name, method name, or specific lines of code targeted for modification, implementing the changes as specified in the action plan. When modifications are suggested for a method, the CodeGeneration Module regenerates the entire method body. Subsequently, the CodeEd- iting Module replaces the original method with this newly generated code, ensuring syntactic correctness and maintaining the overall structure of the codebase. This bifurcated approach allows for a separation of concerns between code generation and code integration, enhancing the system's ability to produce contextually appropriate modifications while preserving the integrity of the existing codebase structure. The use of AST- based editing ensures that the modifications are implemented with precision, reducing the risk of introducing syntactic errors or inconsistencies. By combining LLM-driven code generation with AST-based code editing, this module achieves a balance between creative problem-solving and structural code maintenance, contributing to the overall robustness and reliability of the SuperCoder2.0 system.

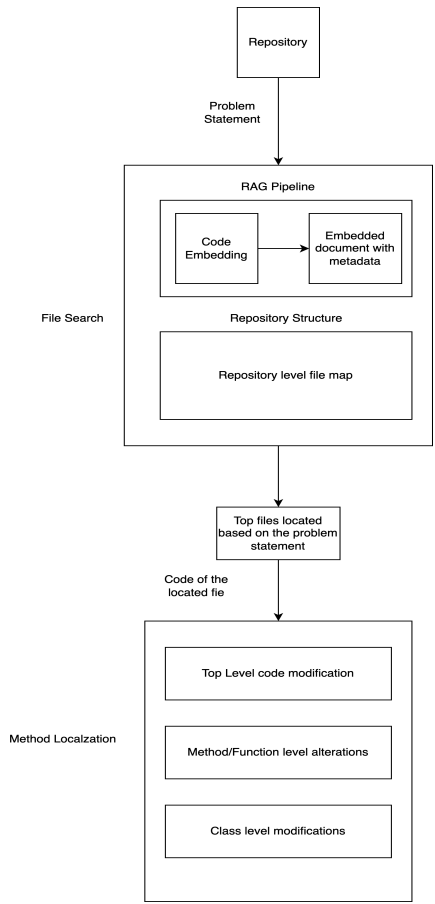

Fig. 1. File and Method localisation

*1) Code Generation and Editing Process:* The code generation and editing process is initiated by invoking an LLM with inputs comprising the action plan, problem statement, and relevant code segments (class, method, or top-level code) identified in the preceding steps. This LLM invocation is executed iteratively for each plan element generated by the CodeParser module. To enhance robustness and solution diversity, we employ a multi-temperature sampling approach. For each plan, 'k' distinct LLM calls are made, each with a different temperature setting, yielding 'k' potential solutions for the given problem statement. This strategy allows for the exploration of various solution spaces, potentially uncovering more optimal or creative resolutions. The editing process is tailored to the hierarchical level of the relevant code location:

- For class-level modifications, the entire class code is replaced with the newly generated class code using AST module.
- For method or function-level changes, the AST module in Python is utilized to replace the entire method or function body.
- For top-level code alterations, the LLM generates start and end line numbers, and the code within this range is substituted with the newly generated code.

Following the generation of 'k' potential solutions, a comprehensive validation process is initiated. Each solution undergoes evaluation against the full suite of repository-level test cases, rather than a subset, ensuring thorough verification of code integrity and functionality. Solutions that induce failures in previously passing tests are systematically eliminated from consideration. In scenarios where multiple viable solutions remain post-validation, an additional LLM call is made. This final invocation is tasked with selecting the optimal solution based on the original problem statement and any additional context derived from the validation process. This multi-stage, temperature-varied generation and rigorous validation approach significantly enhances the reliability and effectiveness of the code modification process. By leveraging both divergent solution generation and convergent solution selection, SuperCoder2.0 maximizes its potential to produce high-quality, contextually appropriate code modifications while minimizing the risk of introducing new errors or regressions.

*C. Iterative Feedback and Refinement Mechanism*

SuperCoder2.0 incorporates a sophisticated feedback and refinement mechanism to ensure the integrity and effectiveness of generated code solutions. This process is structured as follows:

- Pre-Application Baseline: Prior to the application of any generated code, a comprehensive execution of all repository-level test cases is conducted. The results, including both passed and failed tests, are meticulously recorded to establish a baseline performance metric.
- Post-Application Evaluation: Following the implementation of each generated solution, the entire suite of repository-level test cases is re-executed. This step is crucial for identifying any newly introduced failures or regressions.
- Feedback Loop Activation: In the event that new test case failures are detected, a feedback loop is initiated. This loop triggers the CodeGeneration module to refine the specific solution that induced the failure. Notably, this refinement process is targeted and does not involve the generation of multiple candidate solutions through temperature sweeps, as in the initial generation phase.
- Iterative Refinement: The feedback and refinement process is applied iteratively to each potential solution. While the system is designed to support multiple iterations of this process for each solution, we have implemented a constraint limiting it to a single iteration to optimize token usage and computational efficiency.
- Contextual Information Provision: The feedback loop supplies the CodeGeneration LLM with detailed information about the failed test case, including the relevant code segment and the original problem statement. This comprehensive context enables the LLM to perform targeted and informed corrections.

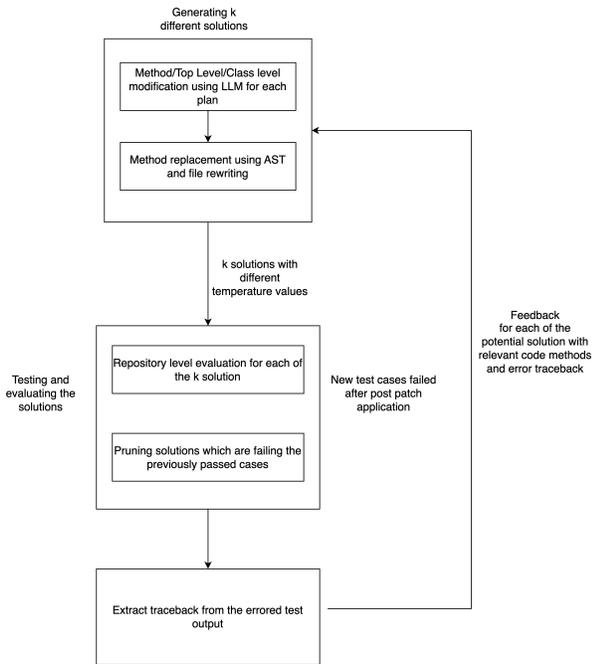

Fig. 2. Code generation and retry mechanism to find viable solution

This iterative feedback mechanism (Figure 2) enhances the system's ability to produce robust, error-free code modifications. By providing specific failure information and allowing for targeted refinements, SuperCoder2.0 can adapt its solutions in response to complex, interdependent code structures and unforeseen edge cases. This approach strikes a balance between solution quality and computational efficiency, contributing to the overall effectiveness and reliability of the autonomous coding system.

## IV. EXPERIMENTAL METHODOLOGY AND DATASET ANALYSIS

### A. Experimental Design on SWE-Bench Lite

The development and experimental evaluation of SuperCoder2.0 were extensively conducted using the SWE-Bench Lite dataset. To ensure rigorous and unbiased assessment, several methodological adjustments were implemented during the evaluation phase:

- Exclusion of Hints: No external hints or guidance were utilized during the problem-solving process.
- Comprehensive Test Suite Execution: Instead of relying on pass-to-pass or fail-to-pass metrics for solution filtering, the entire test suite was executed on the repository both before and after the application of each generated solution. This approach provides a more holistic evaluation of solution impact.
- Targeted Feedback Utilization: Feedback mechanisms were exclusively applied to test cases that transitioned from a passing state to a failing state post-solution implementation. This targeted approach focuses on addressing newly introduced issues.
- Delayed Test Patch Application: The test patch was applied only after the final candidate solution was selected, thereby preventing any potential information leakage that could bias the solution generation process.

### B. Dataset Characteristics and Selection Rationale

The study employs the SWE-bench Lite dataset, a carefully curated subset of the original SWE-bench dataset provided by the SWE-Bench team itself.

*1) Original SWE-bench Dataset:* The SWE-bench dataset was initially conceived to offer a comprehensive array of codebase problems verifiable through in-repository unit tests. It encompasses 2,294 issue-commit pairs distributed across 12 distinct Python repositories. While extensive, the dataset presented significant challenges:

Computational Intensity: The scale of SWE-bench imposed substantial demands on time and computational resources for thorough evaluation. Complexity: The intricate nature of the problems made SWE-bench a formidable benchmark, potentially deterring systems aimed at incremental progress.

*2) SWE-bench Lite: A Focused Subset:* To address these limitations, SWE-bench Lite was developed, comprising 300 carefully selected instances from the original dataset. Key features of this subset include:

Self-Containment: Problems are more self-contained, facilitating focused evaluation. Functional Bug Fix Emphasis: The subset prioritizes the assessment of functional bug-fixing capabilities. Preserved Diversity: SWE-bench Lite maintains the repository distribution and problem diversity of the original dataset. Computational Efficiency: The reduced scale allows for more manageable and efficient computational evaluations.

This strategic subset selection enables a more accessible yet representative benchmark, striking a balance between comprehensive evaluation and practical feasibility. It provides a platform for assessing incremental progress in autonomous coding systems while retaining the challenging and diverse nature of real-world software engineering problems.

### C. Evaluation Metrics and Performance Analysis

To rigorously assess the efficacy of SuperCoder2.0 on the SWE-Bench Lite dataset, we employed a multi-faceted evaluation framework encompassing these primary metrics:

- Resolution Rate: Defined as the ratio of successfully resolved instances to the total number of instances (300) in the dataset. This metric provides a direct measure of the system's problem-solving capability across diverse coding challenges.
- File-Level Localization Accuracy: Measures the system's ability to correctly identify the file(s) containing the bug or requiring modification.

These granular localization metrics provide valuable insights into the system's diagnostic capabilities, which are crucial for efficient problem-solving in large codebases. The combination of these metrics offers a comprehensive view of SuperCoder2.0's performance, balancing between problem-solving effectiveness, computational efficiency, and precision in issue localization. This multi-dimensional evaluation approach allows for a nuanced understanding of the system's strengths and areas for potential improvement, facilitating targeted enhancements in future iterations of the technology.

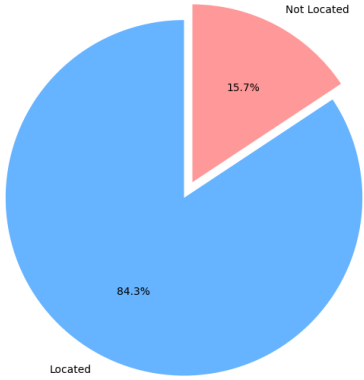

Fig. 3. Percentage of correct file among the top-5 candidate files

| Framework/Tool | Resolved (in %) |
| --- | --- |
| CodeStory Aide + Mixed Models | 43.00 |
| AbanteAI MentatBot + GPT 4o (2024-05-13) | 38.00 |
| Gru(2024-08-11) | 35.67 |
| **SuperCoder2.0** | **34.00** |
| Bytedance MarsCode Agent + GPT 4o (2024-05-13) | 34.00 |
| Alibaba Lingma Agent | 33.00 |
| AutoCodeRover | 30.67 |
| Amazon Q Developer Agent | 29.67 |
| Agentless + RepoGraph + GPT-4o | 29.67 |
| CodeR + GPT 4 | 28.33 |
| MASAI + GPT 4o | 28.00 |
| SIMA + GPT 4o | 27.67 |
| Moatless Tools + Claude 3.5 Sonnet | 26.67 |
| OpenDevin + CodeAct v1.8 | 26.67 |
| IBM Research Agent-101 | 26.67 |
| Aider + GPT 4o & Claude 3 Opus | 26.33 |
| OpenCSG StarShip CodeGenAgent + GPT 4 (0613) | 23.67 |
| SWE-agent + Claude 3.5 Sonnet | 23.00 |

TABLE I
PERFORMANCE OF OTHER MODELS AND FRAMEWORKS ON SWE BENCH LITE

## V. RESULTS AND ANALYSIS ON SWE-BENCH LITE

The empirical evaluation of SuperCoder2.0 on the SWE-Bench Lite dataset yielded several significant insights into its performance and effectiveness in autonomous code modification tasks.

### A. File Localization Efficiency

Our analysis revealed that the combination of Agentless's hierarchical file structure approach with our retrieval-based method resulted in a marginal but notable improvement in performance. We posit that LLMs may have an inherent ability to infer relevant files from file structures, possibly due to exposure to similar patterns during pre-training. However, this hypothesis requires further investigation for conclusive validation. The file localization efficiency was quantitatively assessed using a top-k analysis, with k values of 1 and 5. Notably, for k=5, SuperCoder2.0 achieved an accuracy of 84.33% in identifying the correct file among the candidate files (Figure 3). This high accuracy underscores the effectiveness of our combined RAG and Repository File Level Map approach in narrowing down the search space for relevant code sections.

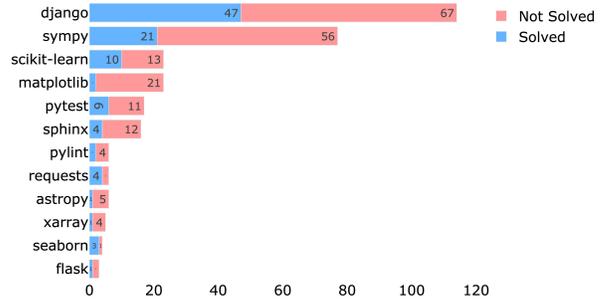

Fig. 4. Repository Wise Performance

### B. Code Analysis and Editing Strategies

Our experiments indicated that providing the entire code file for bug localization was more effective than supplying a high-level map alone. This observation has led to the development and adoption of a hierarchical search space reduction strategy, which has been subsequently implemented in various other works in the field. In the context of code editing, particularly within the SWE-Bench Lite framework, we found that replacing entire method bodies yielded optimal results. This approach significantly reduced indentation errors compared to more granular cut-and-insert techniques. This finding opens avenues for future research into more sophisticated code editing methodologies.

### C. Embedding Model Efficacy

The implementation of Jina Code Embeddings for natural language to code mapping in our RAG system proved highly effective. This success suggests that further development and refinement of embedding models specifically tailored for NLP-to-Code applications could yield substantial improvements in autonomous coding systems.

### D. Cross-Repository Performance

Analysis of SuperCoder2.0's performance across different repositories in the test set (Figure 4) demonstrates its versatility in handling diverse codebases and problem types. This cross-repository consistency is a crucial indicator of the system's robustness and adaptability to varied software engineering challenges.

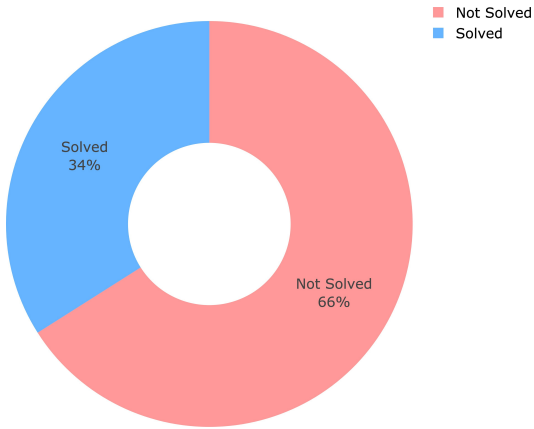

Fig. 5. Repository Wise Performance

### E. Overall Performance Metrics

SuperCoder2.0 successfully resolved 34% of the instances in the SWE-Bench Lite dataset (Figure 5). This performance places the system globally on the SWE-Bench leaderboard at the time of this writing. Table I provides a comparative analysis of SuperCoder2.0's performance against other state-of-the-art models and frameworks.

### F. Comparative Analysis

When compared to other leading systems (Table I), SuperCoder2.0's performance is competitive, particularly considering its novel approach to code localization and editing. The system's performance is comparable to established frameworks like Bytedance MarsCode Agent and surpasses several other prominent systems, including AutoCodeRover and Amazon Q Developer Agent. These results collectively demonstrate SuperCoder2.0's efficacy in autonomous code modification tasks, highlighting its strong performance in file localization and its ability to handle diverse coding challenges across various repositories. The system's competitive standing on the SWE-Bench Lite leaderboard underscores its potential as a robust tool for automated software engineering tasks.

## VI. CONCLUSION AND FUTURE WORK

This paper presents SuperCoder2.0, an advanced autonomous system for software development that demonstrates significant capabilities in code navigation, bug localization, and automated code modification. Through a novel combination of hierarchical search space reduction, retrieval-augmented generation, and structure-aware code editing, SuperCoder2.0 achieves competitive performance on the SWE-Bench Lite dataset, resolving 34% of instances. Key contributions of this work include:

A three-step hierarchical approach to codebase navigation and bug localization, combining RAG and file-level mapping techniques. An innovative code editing strategy that preserves code structure and integrity while implementing comprehensive modifications. An iterative feedback and refinement mechanism that enhances solution quality and reliability.

The system's performance, particularly its 84.33% accuracy in file localization within the top 5 candidates, underscores the effectiveness of our combined RAG and Repository File Level Map approach. Furthermore, SuperCoder2.0's ability to handle diverse repositories and problem types demonstrates its potential as a versatile tool for automated software engineering tasks. Future work will focus on several areas like refining the code editing process to further reduce indentation errors and improve code quality. Exploring advanced embedding models specifically tailored for NLP-to-Code applications to enhance the system's understanding and generation capabili- ties. Investigating the potential for multi-agent architectures to handle more complex software engineering tasks. Extending the system's capabilities to support multiple programming languages and diverse software development paradigms.

As autonomous coding systems continue to evolve, SuperCoder2.0 represents a significant step towards more efficient, reliable, and adaptable software development processes. By addressing the challenges of code navigation, bug localization, and automated code modification, this research contributes to the ongoing advancement of AI-assisted software engineering, paving the way for more sophisticated and capable autonomous programming systems in the future.